\documentclass{aastex}
\usepackage{emulateapj5,psfig}
\usepackage{graphicx}
\usepackage{timesfonts}
\makeatletter
{\end{center}\end{minipage}\smallskip}

\newenvironment{inlinefigure}{%
\def\@captype{figure}%
\noindent\begin{minipage}{0.999\linewidth}\begin{center}}
{\end{center}\end{minipage}\smallskip}
\makeatother
\def\ltsima{$\; \buildrel < \over \sim \;$}
\def\lsim{\lower.5ex\hbox{\ltsima}}
\def\loe{\lower.5ex\hbox{\ltsima}}
\def\gtsima{$\; \buildrel > \over \sim \;$}
\def\gsim{\lower.5ex\hbox{\gtsima}}
\def\goe{\lower.5ex\hbox{\gtsima}}

\newcommand{\s}{$^{s}~$}
\def\uu {4U\,0142$+$614\,}
\def\src {1E\,1048.1$-$5937\,}

\def\axj {AX\,J1844$-$0258\,}

\def\ee {1E\,2259$+$586\,}
\def\s19 {SGR\,1900$+$14\,}
\def\s18 {SGR\,1806$-$20\,}

\def\xte {XTE\,J1810$-$197\,}

\def\ergscm {\rm erg\, cm^{-2}\, s^{-1}}
\newcommand{\CXO}{{\em Chandra}\,}

\newcommand{\xmm}{{\em XMM--Newton}\,}
\newcommand{\BSAX}{{\em Beppo}SAX\,} 
\newcommand{\RXTE}{{\em R}XTE\,}
\newcommand{\bc}{\begin{center}}
\newcommand{\ec}{\end{center}}
\def\ltsima{$\; \buildrel < \over \sim \;$}
\def\lsim{\lower.5ex\hbox{\ltsima}}
\def\loe{\lower.5ex\hbox{\ltsima}}
\def\gtsima{$\; \buildrel > \over \sim \;$}
\def\gsim{\lower.5ex\hbox{\gtsima}}
\def\goe{\lower.5ex\hbox{\gtsima}}
\newcommand {\rc}{\rm}

\lefthead{ISRAEL ET AL.}
\righthead{UNVEILING THE NATURE OF \src}
\submitted{Received: 11 November 2003; ~~ Accepted: 15 January 2004}

\begin{document}

\title{Accurate X--ray position of the Anomalous X--ray Pulsar \xte\
and identification of its likely IR counterpart\footnotemark[1]}
\vspace{-3mm}
\footnotetext[1]{\sc The results reported in this paper are partially based on 
observations carried out at ESO, Cerro Paranal, Chile (072.D-0297)}
\vspace{5mm}

\authoremail{gianluca@mporzio.astro.it}

\author{
G.L. Israel\altaffilmark{2,3}, N. Rea\altaffilmark{4},
V. Mangano\altaffilmark{2}, V. Testa\altaffilmark{2},
R. Perna\altaffilmark{5}, W. Hummel\altaffilmark{6},
R. Mignani\altaffilmark{6}, N. Ageorges\altaffilmark{7}, 
G. Lo Curto\altaffilmark{7}, O. Marco\altaffilmark{7},
L. Angelini\altaffilmark{8}, S. Campana\altaffilmark{9},
S. Covino\altaffilmark{9}, G. Marconi\altaffilmark{7},
S. Mereghetti\altaffilmark{10}, and
L. Stella\altaffilmark{2,3}}


\affil{2. INAF -- Osservatorio Astronomico di Roma, V. Frascati 33, 
       I--00040 Monteporzio Catone (Roma), 
       Italy; gianluca , mangano, testa and stella@mporzio.astro.it}

\affil{3. Affiliated to the International Center for Relativistic 
       Astrophysics}

\affil{4. Universit\'a degli studi di Roma ``Tor Vergata'', Via 
       della Ricerca Scientifica 1, I--00133 Roma, Italy; 
       nandad@mporzio.astro.it }

\affil{5. Department of Astrophysical Sciences, 
       Princeton University, Princeton, NJ 08544-1001, USA; 
       rosalba@astro.princeton.edu}

\affil{6. European Southern Observatory, Karl--Schwarzschildstr. 2, 
       D--85748 Garching, Germany; rmignani and whummel@eso.org}

\affil{7. European Southern Observatory, Casilla 19001, Santiago, 
       Chile; gmarconi, ageorges, amarco and glocurto@eso.org}

\affil{8. Laboratory of High Energy Astrophysics, Code 660.2,
       NASA/Goddard, Space Flight Center, MD 20771, USA; 
       angelini@davide.gsfc.nasa.gov}

\affil{9. INAF -- Osservatorio Astronomico di Brera, Via Bianchi 
       46, I--23807 Merate (Lc), Italy; covino and campana@merate.mi.astro.it}

\affil{10. CNR--IASF, Istituto di Astrofisica Spaziale e Fisica
       Cosmica, Sezione di Milano ''G.Occhialini'', Via Bassini 15, I--20133
       Milano, Italy; sandro@mi.iasf.cnr.it}

\begin{abstract}
{\rc We report the accurate sub-arcsec X--ray position of the new
Anomalous X--ray Pulsar (AXP) \xte, derived with a \CXO--HRC Target of
Opportunity observation carried out in November 2003}. We also report
the discovery of a likely IR counterpart based on a VLT (IR band)
Target of Opportunity observation carried out in October 2003.
Our proposed counterpart is the only IR source ($Ks$=20.8) in the
X--ray error circle. {\rc Its IR colors as well as the X-ray/IR flux
ratio, are consistent with those of the counterparts of all other AXPs
(at variance with field star colors).}  Deep Gunn-$i$ band images
obtained at the 3.6m ESO telescope detected no sources down to a
limiting magnitude of 24.3.
Moreover, we find that the pulsed fraction and count rates of \xte\
remained nearly unchanged since the previous \CXO\ and \xmm\
observations (2003 August 27th and September 8th, respectively). We
briefly discuss the implications of these results. In particular, we
note that the transient (or at least highly variable) nature of this
AXP might imply a relatively large number of hidden members of this
class.
\end{abstract}


\section{INTRODUCTION} 

Despite their limited number, Anomalous X--ray Pulsars (AXPs) are one
of the most intensively studied galactic high energy sources.  {\rc
Since these sources were first proposed as a separate class
(Mereghetti \& Stella 1995, van Paradijs et al. 1995), their number
has grown slowly and there are now five confirmed AXPs plus two
candidates}. These sources are relatively slowly-rotating,
spinning-down, radio-quiet X-ray pulsars with no evident signature for
binary motion, and X-ray luminosity exceeding the rotational energy
loss by a large factor (see Israel, Mereghetti \& Stella 2002 and
Mereghetti et al. 2002 for a review). AXPs have been linked to SGRs
because of similar timing properties, namely spin periods (P in the
5-12s range) and large period derivatives (\.P $\sim 10^{-11}-10^{-12}
{\rm s/s}$).  It is now commonly believed that Soft $\gamma$--ray
Repeaters (SGRs) are magnetars, isolated neutron stars powered by the
decay of theirstrong, super-critical magnetic fields of
B$>$10$^{14}$\,Gauss (Duncan \& Thompson 1992; Thompson \& Duncan
1995; {\rc Kouveliotou et al. 1998}). The same model has been applied
to AXPs. Nonetheless, there is a growing group of radio pulsars
(Camilo et al. 2000; McLaughlin et al. 2003) with comparably long
periods and inferred magnetic field strengths approaching
$10^{14}$\,G.  These radio pulsars possess no special attribute
linking them to either AXPs (no steady bright quiescent X--ray
emission; Pivovaroff, Kaspi \& Camilo 2000) or SGRs (no bursting
episodes).  Thus periodicity alone does not appear to be a sufficient
attribute for classification.  Correspondingly, a very high magnetic
field strength cannot be the sole factor governing whether a neutron
star is a magnetar, a radio pulsar or an accreting neutron star.

The recent detection of X--ray bursts from 1E\,2259+586 and
1E\,1048.1--5937 has strengthened the connection of AXPs with SGRs
(Kaspi \& Gavriil 2002; Gavriil et al. 2002). At the same time, the
study of the former sources resulted in the identification of new
observational properties. In the case of 1E\,2259+584, IR variability
of the counterpart was detected a few days after an episode of strong
X--ray bursting activity (Kaspi et al 2003).  IR variability of the
counterpart to 1E\,1048.1--5937 was also detected; this might be
related to the X--ray variability which was observed from this source
(Israel et al. 2002). These findings have opened a new perspective in
the field, challenging the predictions of current models.  It is yet
clear which specific physical parameter(s) differentiate(s) AXPs from
SGRs (if any).

A pronounced X--ray variability seems to be the main new
characteristic of the recently-proposed member of this class of
pulsars, namely \xte\ (also known as CXOU J180951.1--194351; Gotthelf
et al. 2003a, hereafter G03). The source was discovered with \RXTE\ in
July 2003 at an absorbed flux level of $\sim 5.5\times10^{-11}
\ergscm$ (2--10 keV; $N_H=1\times10^{22}$ atoms\,${\rm cm^{-2}}$;
Markwardt et al. 2003a; Ibrahim et al. 2003).  Subsequent
re-examination of archival data showed that the source was present in
the \RXTE\ PCA data since February 2003 with a flux of $\sim
8.64\times10^{-11} \ergscm $ (2--10 keV).  The source shows a soft
two--component spectrum, pulses at a period P of 5.5\,s and a period
derivative \.P of 1.8$\times$10$^{-11}$\,$ss^{-1}$ (Markwardt et
al. 2003b, Ibrahim et al. 2003, Tiengo \& Mereghetti 2003). No
relatively bright optical ($I>$21.3) and IR ($K=$17.5) counterpart was
found in the 2\farcs5 radius error circle obtained with \CXO\ in
August 2003 (1$\sigma$ confidence level; G03, Ibrahim et al. 2003).

In this paper, we present the results from \CXO\ and VLT Target of
Opportunity observations of \xte\ that we obtained in the fall of
2003. These led to a subarcsec accurate X-ray position for \xte\ and
to the identification of its likely IR counterpart. {\rc Preliminary
results from this study were reported in Israel et al 2003c (posted on
2003 November 4th).}  We found that the optical/IR colours and broad
band spectrum of this pulsar are similar to those of AXPs, strongly
suggesting that \xte\ belongs to the same class.  {\rc Based on these
findings and on the X-ray variability previously reported for this
pulsar (Ibrahim et al. 2003), we conclude that \xte\ represents the
first confirmed AXP showing a transient behavior (in the X-ray band).}

\section{\CXO\ OBSERVATIONS}

The field of \xte\ was observed by \CXO\ with the High Resolution
Imager (HRC--I; Zombeck et al. 1995) on 2003 November 1st for an
effective exposure time of 2866\,s. Data were reduced with {\tt CIAO
3.0.1} and analysed with standard software packages for X--ray data
({\tt Ximage}, {\tt Xronos}, etc.).  Only one source was detected in
the HRC--I (see Israel et al. 2002 for details on the detection
algorithm). The source has the following coordinates: R.A. = 18$^{\rm
h}$09$^{\rm m}$51\fs08, DEC. = --19\arcdeg43\arcmin51\farcs74
(statistical uncertainty of 0\farcs13; equinox J2000), with a total
uncertainty circle radius of 0\farcs7 (90\% confidence level; Israel
et al. 2003c). The position is consistent with that of the previous
\CXO\ observation (G03). Photon arrival times were extracted from a
circular region with a radius of 1\farcs5, including more than 90\% of
the source photons, and corrected to the barycenter of the solar
system. Coherent pulsations at a period of about 5.5\,s were detected,
confirming that the source was indeed \xte. To refine the period
determination we adopted a phase fitting technique. The best pulse
period was determined to be P=5.5391$\pm$0.0004\,s (90\% confidence
level). The pulsed fraction in the HRC--I energy band was 49$\pm$2\%
(semi--amplitude of modulation divided by the mean source count
rate). The latter value is consistent with that reported by G03
(46$\pm$3\% using our pulsed fraction definition).  The source count
rate was 1.04$\pm$0.04\,ct\,s$^{-1}$, marginally lower than the
previous \CXO\ observations carried out 66 days earlier
(0.96$\pm$0.04\,ct\,s$^{-1}$; G03).

Finally, the spatial profile was found to be in good agreement with the
expected \CXO\ Point Spread Function (PSF) for an on--axis source (see
Israel et al. 2002 for details).

\section{OPTICAL/IR OBSERVATIONS}

Data were acquired at VLT-UT4 Yepun with the Nasmyth Adaptive Optics
System and the High Resolution Near IR Camera (NAOS-CONICA) on 2003
October 7th. The pixel size of the camera is 0.027\arcsec\ and the
FWHM is approximately 6 pixels.  Images have been reduced with
instrument-specific pipelines and then co-added. A total of 21 images
(effective exposure time of each frame is 120\,s obtained by averaging
three exposures of 40\,s on the chip) in $H$ and $Ks$ filters were
used for the analysis.  Zero-points were obtained by using the
standard S234-E from the catalog of Persson et al. (1998) obtaining
the following values: ZP($H$)=23.80, ZP($Ks$)=22.91. Thanks to
adaptive optics we obtained an on--axis source PSF of $\sim$0\farcs15
and we inferred a limiting magnitude (S/N$\sim$3) of 22.5 ($H$) and
21.5 ($Ks$).  A preliminary catalog for the objects in the field was
obtained by co-adding $H$ and $Ks$ images.  Magnitudes and positions
of detected stars in the coadded images were derived by means of
aperture photometry and centroid determination. (Stetson 1990).

Optical observations were carried out on 2003 October 12th with EFOSC2
at the 3.6m ESO telescope (La Silla).  The night was clear with
variable seeing (between 0.7" and 1.5").  The \xte\ field was imaged
with 21 exposures of 240s each using the Gunn--$i$ filter (central
wavelength 803.4nm; FWHM of 151nm).  The zero point for the night is
extracted from the tabulated Johnson and Cousins $I$ magnitudes for
the standard field, and its value is: ZP($I$)=23.85 (Landolt 1992).
Data were flat field corrected and fringe pattern subtracted.  Using a
S/N=3 threshold, a PSF area of 11 pixels and the average of the
tabulated extinction values for La Silla, the $I$ limiting magnitude
was 24.30$\pm$0.03.

To register the \CXO\ coordinates of \xte\ on our IR images, we
computed the image astrometry using, as a reference, the positions of
stars selected from the GSC2.2 and 2MASS catalogs which has an
intrinsic absolute accuracy of about 0\farcs2--0\farcs4 (for GSC2.2
and depending on magnitude and sky position of stars).  After taking
into account the uncertainties in the source X--ray coordinates
(0\farcs7), the rms error of our astrometry (0\farcs06) and the
propagation of the intrinsic absolute uncertainties on the catalog
coordinates (we assumed a value of $\sim$0\farcs3), we estimated an
accuracy of about 0\farcs8 to be attached to the \xte\
position. Figure\,1 shows a region of 18\arcsec$\times$18\arcsec\
around the \xte\ position in the $Ks$ band VLT image (the 1$\sigma$
confidence level uncertainty circle reported by G03 is also
superimposed).

\section{DISCUSSION}

The 3.6m ESO data show no object within the new \CXO\ uncertainty
circle down to a $I$ limiting magnitude of 24.3 (S/N$\sim$3).  On the
other hand, only one faint point--like object is present in the VLT
$H$ and $Ks$ frames (marked {\tt X1} in Figure\,1; $H$=22.0$\pm$0.1,
$Ks$=20.8$\pm$0.1).  Note that the probability of finding by chance an
object unrelated to the X--ray source within the \CXO\ uncertainty
region is of the order of $\sim$50\% (number of detected sources
normalised to the area of the \CXO\ error circle). No further object
was detected in the \CXO\ circle down to a limiting $Ks$ magnitude of
about 22.5 (S/N$\sim$1.5). The colors of the IR source, $H-Ks=1.2$ and
$i-Ks>3.5$ are at variance with those of field stars which lie far
from nearby stars in a color-color diagram.  On the contrary the IR
colors are to those of other AXP counterparts.  As an example, the
persistent optical/IR emission of \uu\ has $H-Ks=1.1$ and $I-Ks=3.8$
colours (Hulleman et al. 2000: Israel et al. 2003c), 
\begin{inlinefigure}
\psfig{figure=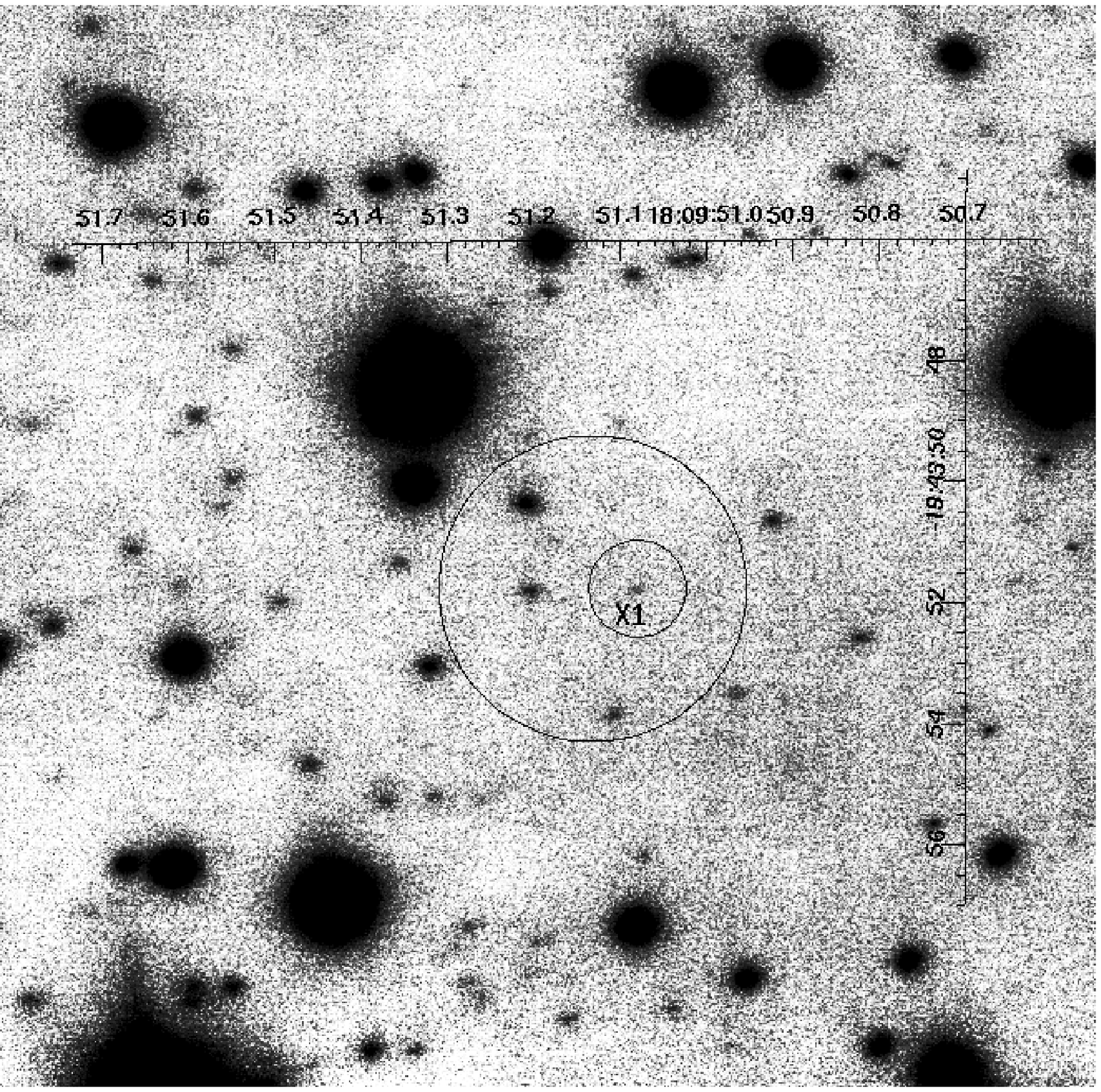,height=9.cm}
\caption{Near--IR $K'$ band image of the region including the position
of \xte, taken with NAOS--CONICA mounted on VLT UT4 Yepun on 2003
October 7th. The two \CXO\ uncertainty circles (2\farcs5 and 0\farcs8
radius at 1$\sigma$ and 90\% confidence level, respectively) are shown.
The proposed IR counterpart is marked with {\tt X1}.  Coordinates are RA
(h\,m\,s) and Dec. ($^o$ \arcmin\ \arcsec; equinox J2000).}
\end{inlinefigure}

\noindent while the ``outbursting'' IR emission of \src\ has
$H-Ks=1.4$ (Wang \& Chakrabarty 2002; note however that the extinction
in the direction of the three objects is different). These findings
make the association of this object with \xte\ quite probable.

In order to further test the hypothesis of the AXP nature of \xte, we
studied the IR-to-X--ray spectrum. We retrieved the \xmm\ archival
data on \xte\ obtained on 2003 September 8th and extracted and fitted
the EPIC--PN spectrum by adopting power-law plus blackbody spectral
model described by G03.  We plotted the IR/optical through X--ray data
of \xte\ in Figure\,2. A value of $A_V$=5.9$\pm$0.3 was used to infer
the unabsorbed IR fluxes and their uncertainties (this was derived
from the N$_H$ inferred from the \xmm\ spectra and $A_V$ = $N_H$ /
(1.79$\times$10$^{21}$\,cm$^{-2}$); Predehl \& Schmitt 1995).  Note
that the count rate, and thus flux, of the second \CXO\ dataset of
\xte\ was nearly unchanged with respect to the previous \CXO\ and
\xmm\ observations; we are thus justified in combining the
IR-to-X--ray measurements plotted in Figure\,2 (we assumed that the
spectral parameters did not change, as suggested also by the constant
pulsed fraction level between the two \CXO\ observations). It is
evident from Figure\,2 that the $F_{\rm X}$ over $F_{\rm IR}$ ratio is
larger than 10$^3$ and similar to those of other AXPs (for a
comparison see Figure\,2 of Israel et al. 2003b; {\rc note however
that the $F_{\rm X}$/$F_{\rm IR}$ ratio is also consistent with that
of low-mass X--ray binaries as in the case of all the other AXPs}).

The relatively high pulsed fraction of \xte\ is not unusual for AXPs:
in fact \src\ has a higher pulsed fraction and additionally, shows
flux variability both in the X--ray and IR bands (Oosterbroek et
al. 1998, Israel et al. 2002).  All the above findings and
similarities with known members of the AXP class, clearly indicate
that \xte\ is an AXP, the one possessing the highest degree of X--ray
flux variability seen so far (a factor of about 100 between quiescent
and outburst peak fluxes). The candidate AXP \axj\ might be another
example of variable/transient AXP (TAXP; Torii et al. 1998; Gotthelf
\& Vasisht 1998). However, \axj\ was caught in a high state only once
and no \.P measurement is available in order to definitively assess
the AXP nature of this source. We note that the quiescent \xmm\ and
\BSAX\ spectrum of \axj\ has a 0.5-10\,keV absorbed flux of
$\sim$3$\times$10$^{-13}\ergscm$ (Israel et al 2003b), quite similar
to that of \xte\ as seen by ROSAT in 1993
($\sim$5$\times$10$^{-13}\ergscm$; G03). Additionally, as in the case
of \xte, also for \axj\ no pulsations were detected in the quiescent
phase (although poor statistics prevented to set any sensitive upper
limit; Israel et al. 2003d).

Regardless of whether or not \axj\ is an AXP, the existence of at
least one TAXP clearly points to a larger number of hidden members of
the AXP class in the Galaxy.  As suggested by G03, part of them might
be the radio-quite X--ray unpulsed Central Compact Objects (CCOs)
found in an increasing number of SNRs. Other AXPs might spend a large
fraction of the time in a quiescent state, and therefore might remain
unidentified as AXPs. There are at least two important new facts that
should be taken into account in the comparison with models: (i) the
X--ray flux variability of more than two orders of magnitude, and (ii)
the non-detection of X--ray pulsations in the quiescent state of \xte\
during a 1996 ROSAT observation.  Variations in the persistent X--ray
emission are common in neutron stars accreting from a companion (White
et al. 1995). The IR measurements presented in this paper for \xte\
rule out any hypothetical main sequence companion star from O to F
spectral-types, and are comparable to those set for other, more
``standard'', AXPs (Mereghetti et al. 1998; Wilson et al 1999).
However, a lighter companion cannot be ruled out (as in the case of
all the other AXPs) and would imply an extremely small and virtually
undetectable Doppler shift in the pulsations (similar e.g. to the 42
minutes orbital period binary system hosting the 7s pulsar
4U\,1626--67; Chakrabarty 1998).

If the two transient objects discovered so far do belong to the same
class of SGRs and AXPs, then one would like to interpret the
observations within the context of the magnetar model.  The magnetar
model as currently formulated does not make specific predictions for
on/off states of the pulsars as well as for their IR
fluxes. Variations in the quiescent emission have been observed
following an X--ray burst in \ee\ (Woods et al. 2003). Soon after the
burst, the persistent X--ray flux was {\rc a factor of about 20
higher}, the temperature also higher, and the blackbody radius much
smaller than in the low quiescent state. Some of the \ee\ properties
in the quiescent emission before and after the burst are consistent
(at least qualitatively), with the those of \xte\ in its low and high
state. Moreover, the slow decay of the ``high-state'' X--ray emission
of \xte\ is found to be in the range of those of SGR 1627-41 and SGR
1900+14 (Ibrahim et al. 2003; {\rc Kouveliotou et al. 2003) for a
power--law decay; therefore we cannot rule out that \xte\ is a SGR. In
this respect, we note that if \xte\ were proven to be a SGR our
proposed IR candidate would be the first ever for a SGR.}

If \xte\ were an isolated object, then (a hypothetical) accretion
would have to proceed through a fall-back disk (Chatterjee et
al. 2000, Alpar 2001). While short-term, small scale fluctuations are
expected in this case, by analogy 
\begin{inlinefigure}
\psfig{figure=f2e.eps,height=9.cm}
\caption{Broad band energy spectrum of \xte.  X--ray raw data are
taken from the EPIC--PN instrument on board \xmm\ (filled squares)
while the solid upper curves are the unabsorbed fluxes for the black
body (BB), the power law (PL), and the sum of the two components. On
the lower left corner of the plot are the optical/IR fluxes/upper
limits: lower points are absorbed values, while upper points are
unabsorbed values for $A_{\rm V}$=5.9.}
\end{inlinefigure}

\noindent with most accreting neutron star systems, large scale
variations might require special conditions. In particular, as the
object spins down, it might occasionally switch from a propeller phase
to an accretion phase. During the propeller phase, accretion is
inhibited, and the star should be bright in X--rays due to its thermal
emission. Indeed, in its low state, \xte\ had an X--ray luminosity of
the order of a few $\times 10^{33} d_5$ erg/s, which is typical of
other known thermal emitting sources (e.g. Becker \& Tr\"umper 1997).
Moreover, in its low state the pulsed fraction of \xte\ (if any)
decreased considerably, suggesting emission from the whole surface of
the star as expected in a cooling object (G03). Note that in a more
general scenario, the CCOs would be AXPs in the above described
state. If the high state is due to resumed accretion onto the magnetic
poles, then larger pulsed fractions would be naturally explained. The
accretion luminosity could be much higher than the thermal emission,
and would dominate the emitted spectrum. These properties appear
consistent with those of \xte, and the properties of the IR
counterpart could be explained in terms of a fall-back disk of size
$\la$ a few $\times 10^{10}$ cm (using the spectral models of Perna et
al. 2000 and Perna \& Hernquist 2000). However, the X--ray flux decay
reported by Ibrahim et al. (2003) would be hardly accounted by the
fall-back model unless the accretion rate were decreasing very
rapidly. Moreover, the above active phase is expected to occur just
once in the AXP life (with a duration longer than the about one year
observed in \xte\ and possibly \axj: Chatterjee et al. 2000).

The pronounced long-term X-ray flux and pulsed fraction variability 
of \xte\ might be more easily explained in the framework of a pulsar in
a binary system with a light companion. In fact, we note that the
quiescent luminosity of \xte\ (and \axj) is similar to that already
observed from transient binary system pulsars (Campana et al. 2002),
together with the pulsed fraction decrease in the on/off transitions
(Campana et al. 2001).  However, in the binary scenario, the
short X--ray bursts displayed by AXPs and, especially, SGRs would be 
difficult to explain (see Mereghetti et al. 2002 for more details).

\acknowledgements

We thank Harvey Tananbaum for making the \CXO\ HRC--I observation
possible through the \CXO\ Director's Discretionary Time program, and
Michael Juda for help during the \CXO\ scheduling phase. We also thank
the ESO Director's Discretionary Time Committee for accepting to
extend our original program allowing us to trigger the VLT
observations, and Lowell Tacconi-Garman for his help and patience
during the OB submission phase. This work is partially supported
through ASI and Ministero dell'Universit\`a e Ricerca Scientifica e
Tecnologica (MURST--COFIN) grants.

\end{document}